# Electron microscopy of microwave-synthesized rare-earth chromites

**Rainer Schmidt**[*,1], **Jesús Prado-Gonjal**[2], **David Ávila**[2], **Ulises Amador**[3] **and Emilio Morán**[2]

[1] *GFMC, Dpto. Física Aplicada III, Facultad de CC. Físicas, Universidad Complutense de Madrid, 28040 Madrid (Spain)*

[2] *Dpto. Química Inorgánica, Facultad de CC. Químicas, Universidad Complutense de Madrid, 28040 Madrid (Spain)*

[3] *Dpto. Química, Facultad de Farmacia, Universidad San Pablo - CEU, 28668 Boadilla del Monte - Madrid (Spain)*

[*] Corresponding author's Email: rainerxschmidt@googlemail.com

**Abstract.** The perovskite rare-earth (RE) chromite series (RE)$CrO_3$ (RE = La, Ce, Pr, Nd, Sm, Eu, Gd, Tb, Dy, Y, Ho, Er, Tm, Yb, Lu) has been synthesized in our laboratory using microwave techniques. In this work we will demonstrate how X-ray diffraction (XRD), Rietveld refinement of XRD pattern and complementary High Resolution Transmission Electron Microscopy (HRTEM) were used to confirm that the desired crystal structure had been formed. Field-emission scanning electron microscopy (FE-SEM) gave clear indications of the shapes and sizes of the powder crystallites. Magnetization measurement data was found to correlate with the structural information obtained from XRD, Rietveld refinement and HRTEM.

## 1. Introduction

### 1.1 Microwave synthesis

The use of microwave irradiation is a promising alternative heat source for the synthesis of inorganic materials. The considerable reduction in reaction temperature and time involved with this recent innovative synthesis technique has the interesting effect that the particle growth during reaction and the resulting particle size of the synthesized products are usually reduced efficiently. In fact, the method is highly sensitive to small changes in the synthesis conditions far away from thermal equilibrium and a fascinating range of different particle shapes and sizes can be obtained. This is not only interesting for potential new applications of nano-sized materials due to novel functionalities, but also facilitates studying fundamental aspects of the physics of condensed matter at the nano-scale. Significant qualitative drawbacks in microwave synthesized materials in terms of crystal quality and physical properties are usually not encountered. In some cases even quantitative improvements, novel crystal arrangements and interesting particle shapes can be achieved. The technique can be regarded a "Fast Chemistry" method. Furthermore, taking into account the low energy requirements, the technique is also consistent with the "Green Chemistry" principles[1]: use of non-hazardous reactants and solvents, high efficiency in terms of energy consumption vs. yield, and easy monitoring to prevent pollution.

Microwave synthesis can be applied to fabricate a large range of technologically important functional materials [2-5]. In each case it is essential to validate that the desired chemical phase has been fully formed, to investigate its microstructure and test the functionality of the product. This requires a set of experimental characterization techniques, where several electron microscopy related methods play a fundamental role. Scanning Electron Microscopy (SEM) gives unambiguous information about the shape of the micro- or nano-sized objects obtained by microwave synthesis. High Resolution Transmission Electron Microscopy (HRTEM) can be used to confirm the crystal structure in real space, determined previously by X-ray diffraction (XRD) and Rietveld refinement of the resulting XRD pattern. Crystal defects can be detected and classified readily Furthermore, the chemical composition of the cation sublattice is accessible by using X-ray energy detectors attached to the microscopes. Such comprehensive analysis allows correlating certain physical properties such as the magnetization behavior to the crystal structure. In most inorganic synthesis research one of the main aims is to establish the relationships between the synthesis procedure, the crystal structure and physical properties. The analysis of these synthesis – structure – property relationships is demonstrated here on the example of the rare-earth (RE) chromites series (RE-$CrO_3$: RE = La, Pr, Nd, Sm, Eu, Gd, Tb, Dy, Y, Ho, Er, Tm, Yb, Lu) [6]. Powders were fabricated by the solid state microwave synthesis method and were investigated by XRD, Rietveld refinement, SEM, HRTEM and magnetization measurements.

1.2 Rare-earth chromites

Rare-earth (RE) orthochromites (RE)$CrO_3$ crystallize in a orthorhombic distorted (Pnma) perovskite structure [7, 8], where the perovskite is probably the most studied structure in materials science [9]. The mineral perovskite is $CaTiO_3$, although the ideal, aristotype structure (the highest symmetry, *Pm-3m*, #221 cubic form) is adopted by $SrTiO_3$ under ambient conditions. The perovskite structure can be described as A cations surrounded by twelve anions (usually oxygen) in cubo-octahedral coordination and B cations surrounded by six anions in octahedral coordination ($ABO_3$). The anions are coordinated by two B-site cations and four A-site cations (Figure 1). The large A cations and O-anions are cubic close packed, with the smaller B cations occupying the octahedral holes between the O-anions.

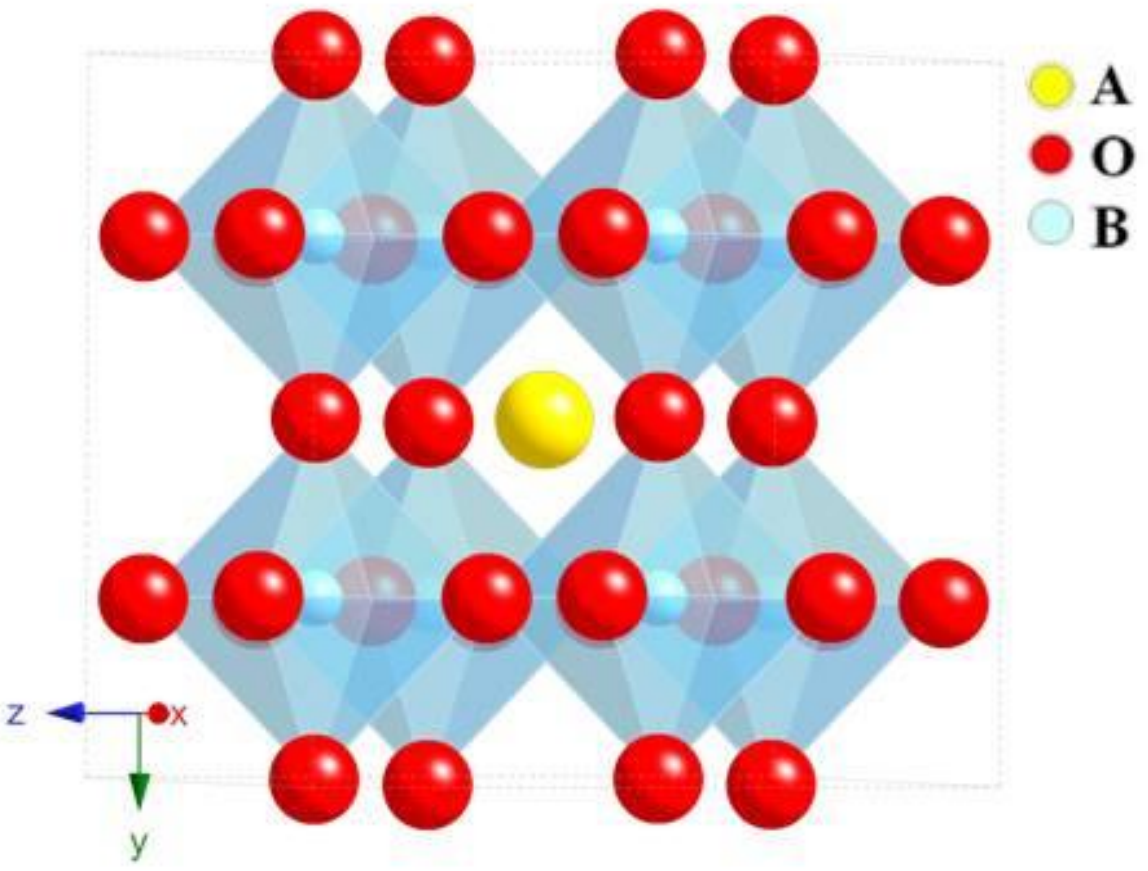

**Fig. 1** 8 unit cells of the ideal cubic $ABO_3$ perovskite structure showing the 12-fold coordination of the A-site cation and the 8-fold octahedral coordination of the B-site cation.

The structure exhibits some flexibility and can accommodate smaller A-site cations than might be expected on the basis of tabulated radii: the B–O–B bond angles can be within a certain range, giving concerted tilting patterns of the ($BO_6$) octahedra to bring oxide ions closer to the A site [10]. This degree of distortion may be predicted to some extent using simple geometric parameters, such as the Goldschmidt tolerance factor (*t*) [11]:

$$t = (r_A + r_O) / \sqrt{2}(r_B + r_O) \qquad (1)$$

where $r_A$, $r_B$ and $r_O$ are the ionic radii of the respective ions. The ideal cubic structure has $t = 1$. This is the case for $SrTiO_3$, where $r_A = 1.44$ Å, $r_B = 0.605$ Å and $r_O = 1.40$ Å. If the A cation is smaller, the tolerance factor is smaller than 1 and the [$BO_6$] octahedra will tilt in order to fill space. The bonding angles of the B-O-B bond will decrease from the ideal 180° to lower values. The cubic perovskite structure occurs for $0.9 < t < 1$. Lower values of $0.7 < t < 0.9$ will lower the symmetry and orthorhombic or rhombohedral perovskite structures are commonly observed. For smaller values of $t < 0.7$ the crystal structure is usually not a perovskite anymore and e.g. the ilmenite structure may occur. On the other hand, if the value is larger $t > 1$, hexagonal variants of the perovskite structure are stable [12, 13].
The octahedral B-O-B bonding angle is well-known to significantly alter the magnetic properties in perovskite oxides, where the magnetic exchange interactions between the electron spins of two magnetic B-site cations are mediated by the oxygen orbitals, which is referred to as a superexchange interaction. The magnetic properties of the orthorhombic (RE)$CrO_3$ perovskite series contain a rich variety of different magnetic spin interactions, where three different types may occur between different types of cation: (1) $Cr^{3+}$-$Cr^{3+}$, (2) $Cr^{3+}$-$(RE)^{3+}$, and (3) $(RE)^{3+}$-$(RE)^{3+}$, with isotropic, symmetric, and antisymmetric anisotropic exchange interactions, respectively [14, 15]. A weak ferrimagnetic moment in some of the species was claimed to arise from Dzyaloshinskii–Moriya (D–M) interactions between $Cr^{3+}$ spins. The antiferromagnetic Neel temperature $T_{N1}$ for $Cr^{3+}$-$Cr^{3+}$ ordering increases with the RE cationic radius, which is associated with diminishing lattice distortions and the reduced $Cr^{3+}$-$O^{2-}$-$Cr^{3+}$ tilting angles [16, 17].
Scientific interest into the RE chromite species is debited to their potential application as multifunctional materials and a possible magneto-electric coupling effect [18, 19]. Initial studies had claimed that certain rare earth chromites belong to a new family of ferroelectric and antiferromagnetic multiferroics [20]. Such claims are somewhat surprising though, because (RE)$CrO_3$ chromites adopt a centrosymmetric orthorhombic structure with space group #62 *Pnma* (or the non-standard setting *Pbnm*). Ferroelectricity in centrosymmetric space groups is unusual but could be explained by Cr off-centering with a local character and small value of displacement leading to weak polarization. The charge transport properties in (RE)$CrO_3$ materials have been claimed to involve p-type semiconductivity with a sensitivity towards humidity, methanol, ethanol, and several gases, which is useful for potential sensor application [21, 22]. Furthermore, $LaCrO_3$ and its doped variants are candidates for application as interconnect materials in solid oxide fuel cells and as catalysts for hydrocarbon oxidation [23].

## 2. Experimental procedure

### 2.1 Microwave synthesis

Precursor nitrates ($Cr(NO_3)_3{\cdot}9H_2O$ 98% and $(RE)(NO_3)_3{\cdot}xH_2O$ 99.9%, Sigma-Aldrich) were mixed with 5% (wt.) amorphous carbon acting as microwave absorber and the mixture was homogenized and compacted into pellets of 12 mm diameter. The pellets were placed in a porcelain crucible and irradiated in a domestic microwave for 10 minutes (2.45 GHz, 800 W). The pellets were then crushed and the amorphous powders heated in a conventional furnace in air at 500 ºC for 2 hours for crystallization to form the precursor $(RE)CrO_4$. Such precursors were then heated at 800 ºC for 2 hours to obtain $(RE)CrO_3$. The resulting powders were compacted again into pellets in a 1 ton die press for 5 minutes and densification sintering was performed in air at 1300º C for 15 hours. The synthesis route is illustrated in Figure 2.

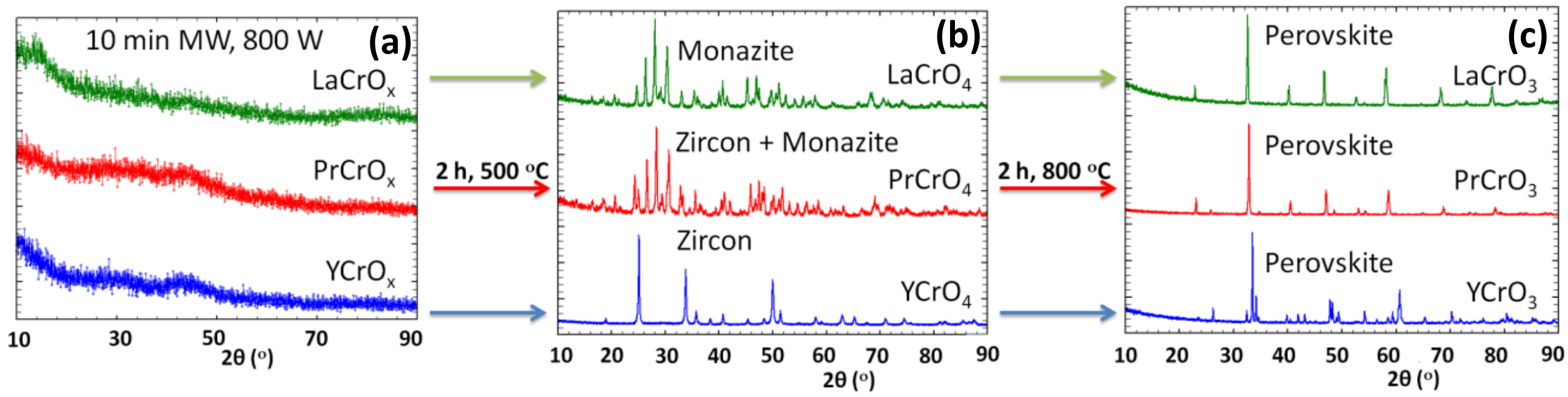

**Fig. 2 (a)** Powder XRD pattern of the MW as-synthesized $LaCrO_x$, $PrCrO_x$ and $YCrO_x$ powders (amorphous), **(b)** XRD pattern of the 500 °C annealed powders showing monazite ($LaCrO_4$), mixed monazite and zircon ($PrCrO_4$), and pure zircon ($YCrO_4$) structure, **(c)** XRD pattern of the 800 °C annealed powders all showing pure perovskite phase ($LaCrO_3$, $PrCrO_3$, $YCrO_3$). Reproduced from Ref.[6] with permission from the American Chemical Society (ACS © 2013).

The reaction mechanism for $(RE)CrO_3$ formation from nitrate precursors can be described by the three-step process indicated by the XRD pattern in Fig.2: (a) Formation of an amorphous phase after 10 min microwave irradiation. (b) Formation of polycrystalline $(RE)CrO_4$ at 500 °C (2h) with zircon-type structure and tetragonal symmetry, S.G. $I4_1/amd$ (# 141), except in the case of $LaCrO_4$ and $PrCrO_4$. In $LaCrO_4$ the monazite-type structure and monoclinic symmetry, S.G. $P2_1/n$ (#14), is obtained due to the large $La^{3+}$ cationic radius, whereas in $PrCrO_4$ zircon and monazite phases coexist. (c) Formation of orthorhombic perovskites $(RE)CrO_3$ at 800 ºC (2 h) with *Pnma* (S.G. #62) symmetry,.

### 2.1 Characterization methods

The phase composition and purity of all precursors, intermediate phases and the final products were tested by powder X-ray diffraction (XRD) on a Bruker D8 high-resolution diffractometer using monochromatic $CuK_{\alpha1}$ ($\lambda$ = 1.5406 Å) radiation obtained from a germanium primary monochromator. X-Rays were detected with a solid-state rapid LynxEye detector. Rietveld refinements of all $(RE)CrO_3$ powder XRD patterns were carried out using FullProf software.
Samples for transmission electron microscopy (TEM) were prepared from powders suspended and ultrasonically dispersed in butanol. One drop of the suspension was placed on a Cu grid with holey carbon film. Selected area electron diffraction (SAED) and high resolution TEM (HRTEM) experiments were performed using a JEOL 3000F microscope with a resolution limit of ≈ 1.1 Å, which is equipped with an X-ray energy dispersive spectroscopy (XEDS) microanalysis system (OXFORD INCA). HRTEM images were recorded with an objective aperture of 70 µm centered on a sample spot within the diffraction pattern area. Fast Fourier Transforms (FFT) of the HREM images were carried out to reveal the periodic contents of the HREM images using the Digital Micrograph package. The experimental HRTEM images were compared to simulated images using MacTempas software. Such computations were performed using information from (a) the structural parameters obtained from Rietveld refinement, (b) the microscope parameters of operating voltage (300 kV) and spherical aberration coefficient (0.6 mm), and (c) the specimen parameters of zone axis and thickness. The defocus ($f$) and sample thickness ($t$) parameters were optimized by assessing the agreement between model and data, leading to values of $f$ = -33 nm and $t$ = 5nm. Field emission scanning electron microscopy (FE-SEM) of Au coated powders was carried out using a Jeol 6335F microscope equipped with a detector for energy-dispersive analysis of X-rays (EDAX). Good homogeneity and the expected semi-quantitative 1:1 cation compositions of RE:Cr for all powders were confirmed by EDAX analysis performed on 10 micro-crystallites (grains), where the cationic ratios on three different areas of each crystallite were measured and the overall average was calculated.
Magnetization measurements of synthesized powders were performed in a Quantum Design XL-SQUID magnetometer in the temperature range of 2 – 300K at 1 kOe. The temperature dependence of the magnetic susceptibility was measured following Zero Field Cooled and Field Cooled (ZFC-FC) procedures with intermediate demagnetization at room temperature. Ferri-magnetic hysteresis was recorded at 10 K from magnetization vs applied field measurements.

## 3. Experimental results

### 3.1 X-ray diffraction and Rietveld refinement

Figure 3 shows the powder X-ray diffraction (XRD) patterns of the (RE)$CrO_3$ series. Advancing in the RE series from lower to higher atomic numbers the cation radius reduces consistently, which leads to a smaller unit cell size of the respective specie in the (RE)$CrO_3$ series. In the XRD pattern the decrease in the unit cell size is obvious from a continuous shift of the intensity peaks to higher angles, since the condition for an XRD intensity maximum shows a reciprocal dependence on the lattice plane spacing of the respective diffraction planes. The diffraction peaks shift to higher angles more clearly for the higher diffraction angles 2-theta, which are more sensitive to changes in the diffraction plane spacing. The diffraction pattern of the $YCrO_3$ specie is included in Figure 3 according to the cationic radius of $Y^{3+}$ in between the $Dy^{3+}$ and $Ho^{3+}$ radii. All XRD pattern in Figure 3 indicate a single phase composition of the powders and Rietveld refinement of all patterns was carried out.

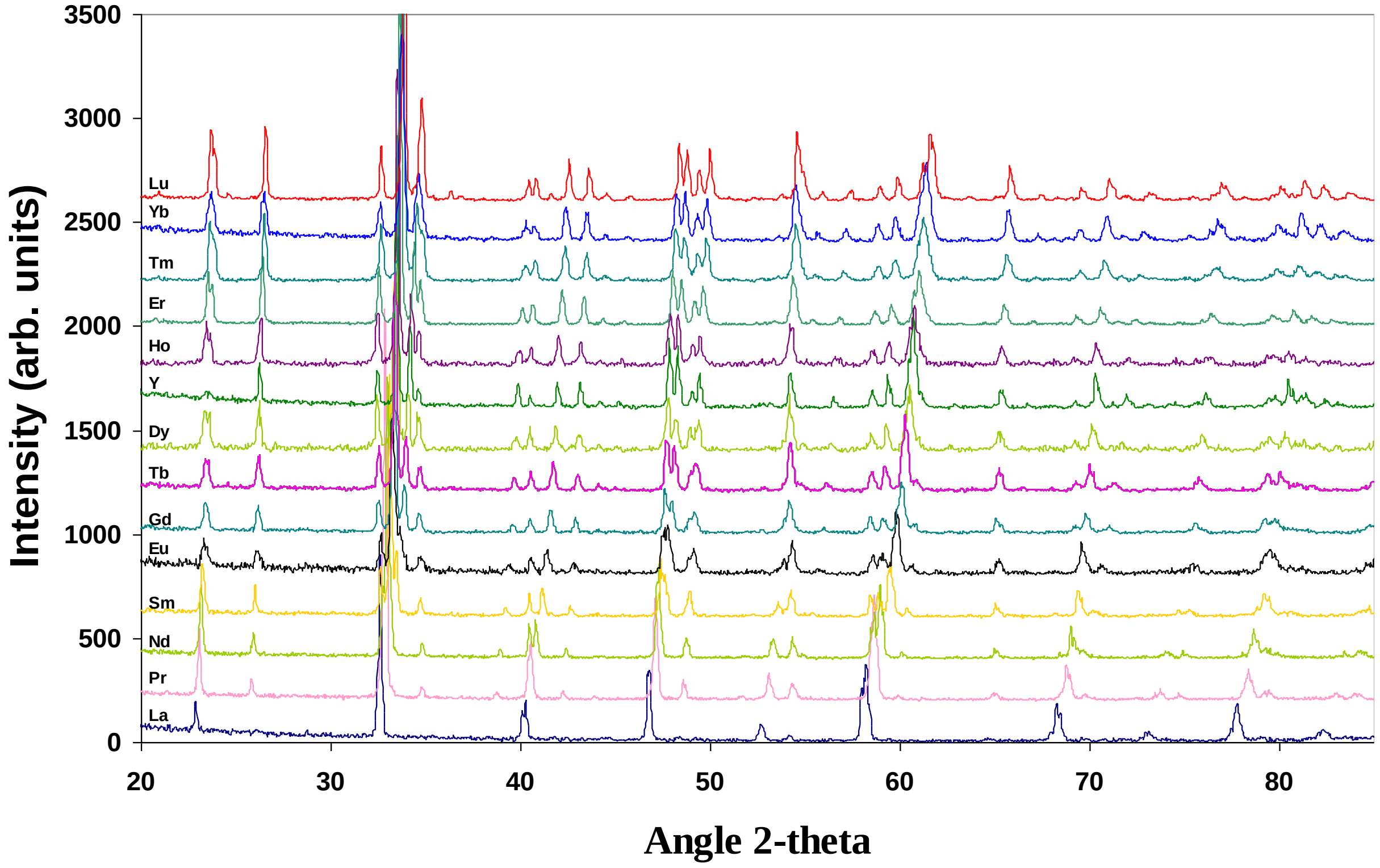

**Fig. 3** XRD patterns of (RE)$CrO_3$ powders. The diffraction peaks move to higher angles with decreasing lattice plane spacing.

The process of Rietveld refinement consists in a fit of the full XRD pattern to a crystal structure model, where the space-group needs to be set and several parameters such as the unit cell vectors, octahedral tilting angles, atomic positions and thermal factors are optimized by the fitting software. The distorted perovskite structure with orthorhombic symmetry (S.G. *Pnma*, #62) was confirmed in all RE-chromites, in agreement with previous work. The perovskite tolerance factors for (RE)$CrO_3$ are $t = 0.85 – 0.93$, which is just about within the reported limits of perovskite structures with orthorhombic symmetry as mentioned above ($0.9 > t > 0.7$). The unit cell parameters $\boldsymbol{a}$, $\boldsymbol{b}$ and $\boldsymbol{c}$ for the RE family obtained from the Rietvelf refinement fits are illustrated in Figure 4a as a function of the RE ionic radius (IOR) and the tolerance factor *t*. IOR values were taken from Shannon's tables [24], where the coordination number 9 was considered for all RE cations consistently.

Figure 4a shows approximately constant cell parameter $\boldsymbol{a}$, whereas $\boldsymbol{b}$ and $\boldsymbol{c}$ increase with increasing IOR. All other parameters obtained from the Rietveld refinement are not presented here. The $Cr^{3+}$-$O^{2-}$ bonding distances were determined with relatively low resolution, but all these distances appear to be similar (≈ 1.97 Å) and close to the expected value (1.9805 Å) [25]. The main effect of the RE size is observed in the octahedral tilting of the perovskite cell. $\theta$, $\varphi$, and $\mu$ (Figure 4b,c) represent the tilting angles of the $BO_6$ octahedra along the three main directions of the ideal $ABO_3$-cubic perovskite structure, which implies that the larger these angles the more distorted the structure. All angles increase quite consistently away from the ideal 0° (undistorted perovskite) with decreasing IOR and increasing RE atomic number, because the rigid $CrO_6$ octahedra accommodate A-site RE cations of decreasing size by decreasing unit cell size and increasing degrees of octahedral tilting. Note that $\theta$ and $\varphi$ were set equal due to symmetry restrictions.

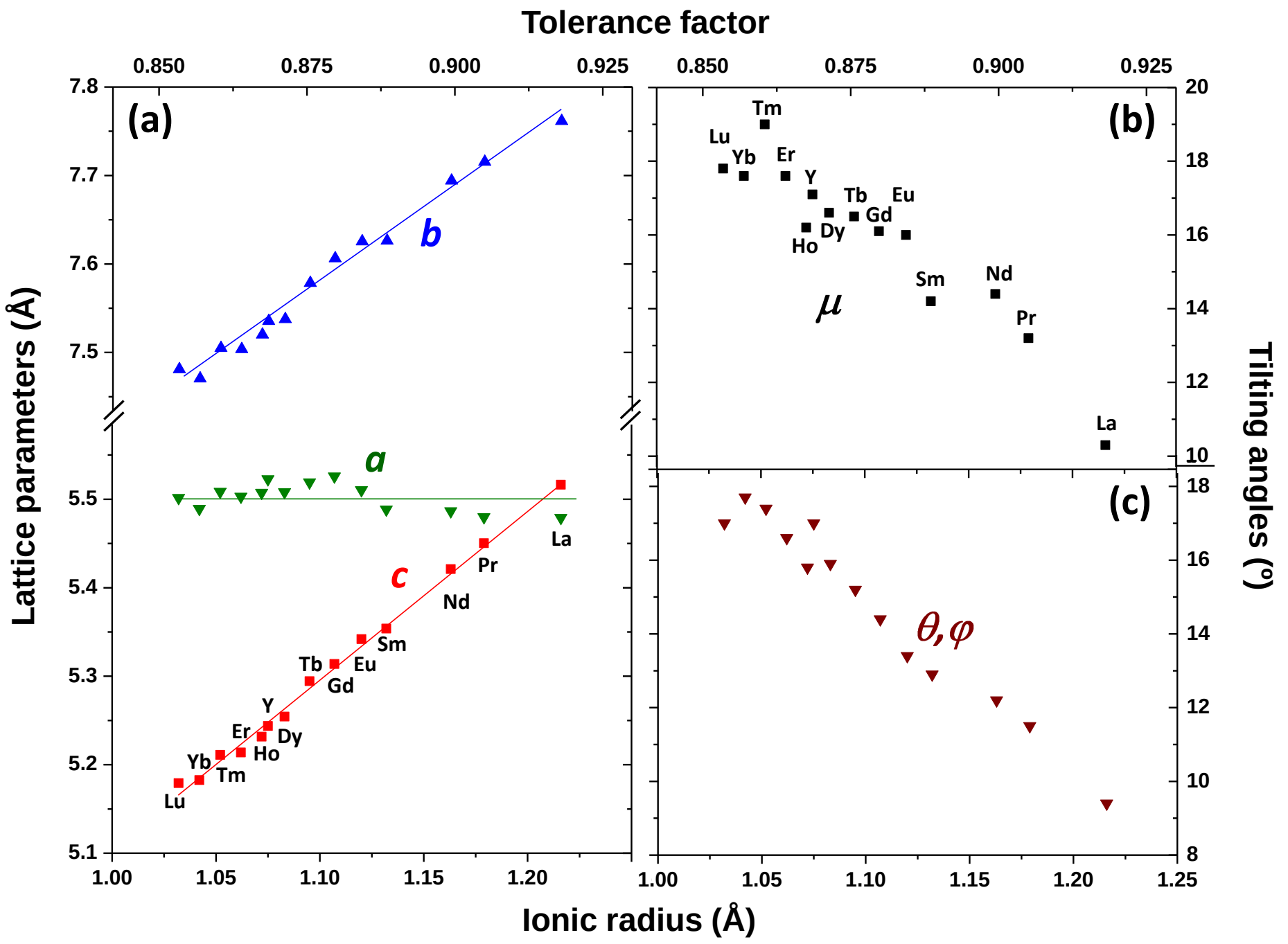

**Fig. 4** **(a)** Lattice parameters ***a***, ***b*** and ***c*** vs the RE ionic radius and the perovskite tolerance factor. The increase of ***b*** and ***c*** with IOR implies an increasing unit cell size. **(b,c)** The decrease of $\theta$, $\varphi$ and $\mu$ with IOR implies decreasing tilting of the perovskite octahedra.

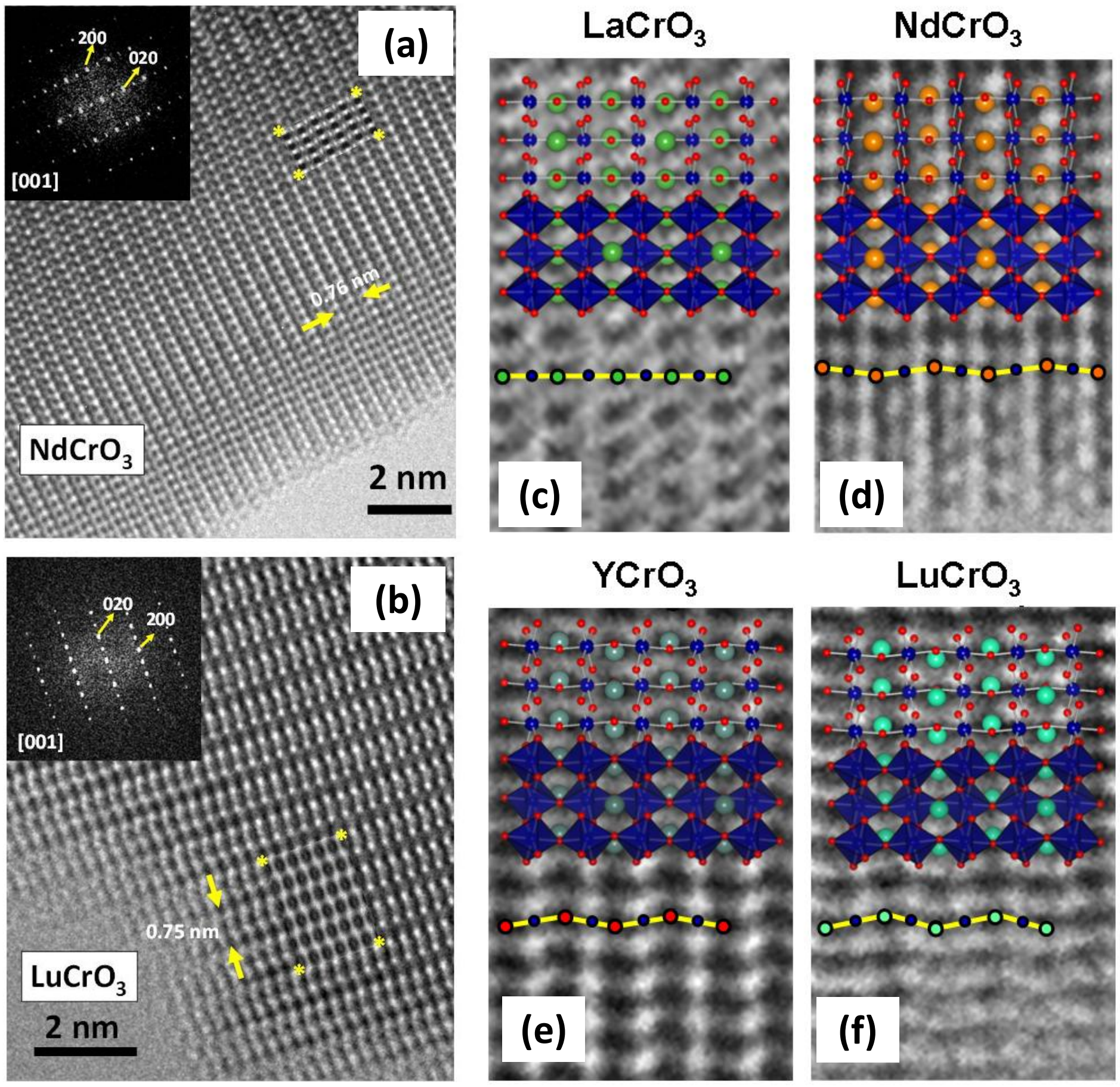

**Fig. 5** **(a,b)** Experimental HRTEM micrograph along the [001] zone axis. No streaking or extra spots are evident in the FFT images in the figure insets. Good agreement between the experimental and simulated images (rectangles marked by yellow asterisks) is demonstrated. **(c-f)** Magnified HRTEM micrographs along the [001] zone axis. The increasing octahedral tilting with decreasing IOR is evidenced by stronger zig-zag orientation of the atoms along the [010] direction. Reproduced from Ref.[6] with permission from the American Chemical Society (ACS © 2013).

3.2 High resolution transmission electron microscopy (HRTEM)

Figures 5a & b show HRTEM images for (RE)$CrO_3$ (RE = Nd and Lu) along the [001] zone axis. The experimental images taken along the [001] zone axis show excellent agreement with the calculated images depicted in the yellow rectangles. In all images the black dots are the projections of the atomic columns from the cationic RE and Cr sublattices, showing a well ordered crystal without the formation of superstructures or defects such as dislocations and stacking faults. This is evidenced by a regular contrast across the HRTEM images and the absence of extra spots or streaking lines in the FFTs (Figure 5a & b insets). An enlargement of the HRTEM images with the projected model from each specie set on top of the experimental image is displayed in Figures 5 c-f. By carefully analyzing the contrast of the images it can be noticed that the darker dots (RE) and the less darker ones (Cr columns) are oriented in zigzag fashion along the [010] direction with increasing amplitude from $LaCrO_3$ < $NdCrO_3$ < $YCrO_3$ < $LuCrO_3$. This feature is a direct result of the octahedral tilting to accommodate RE cations of decreasing size into the octahedral framework. It can be concluded that HRTEM is a powerful technique to display the octahedral tilting in the (RE)$CrO_3$ series in real space, which directly confirms the structural analysis by XRD and Rietveld refinement. The RE:Cr cation ratio in all species was measured by XEDS on different points of more than 20 single crystallites and shows good agreement with the nominal composition. These results confirm the EDAX analyses mentioned above.

3.3 Field-emission scanning electron microscopy (FE-SEM)

Figures 6 a-n-display a full set of SEM micrographs for (RE)$CrO_3$ powders, where clear evidence for agglomeration of sub-micrometer particles is demonstrated. Apparently, the different RE-cations have a significant influence on the synthesis kinetics, leading to a large variety of different particle shapes and sizes, although the synthesis process was equivalent for all (RE)$CrO_3$ species. Figure 7 o shows a $ErCrO_3$ pellet [26], which could be densified efficiently during sintering. It can be concluded that FE-SEM is a powerful technique to display the morphology of the microwave synthesized (RE)$CrO_3$ powders and sintered pellets in very good detail.

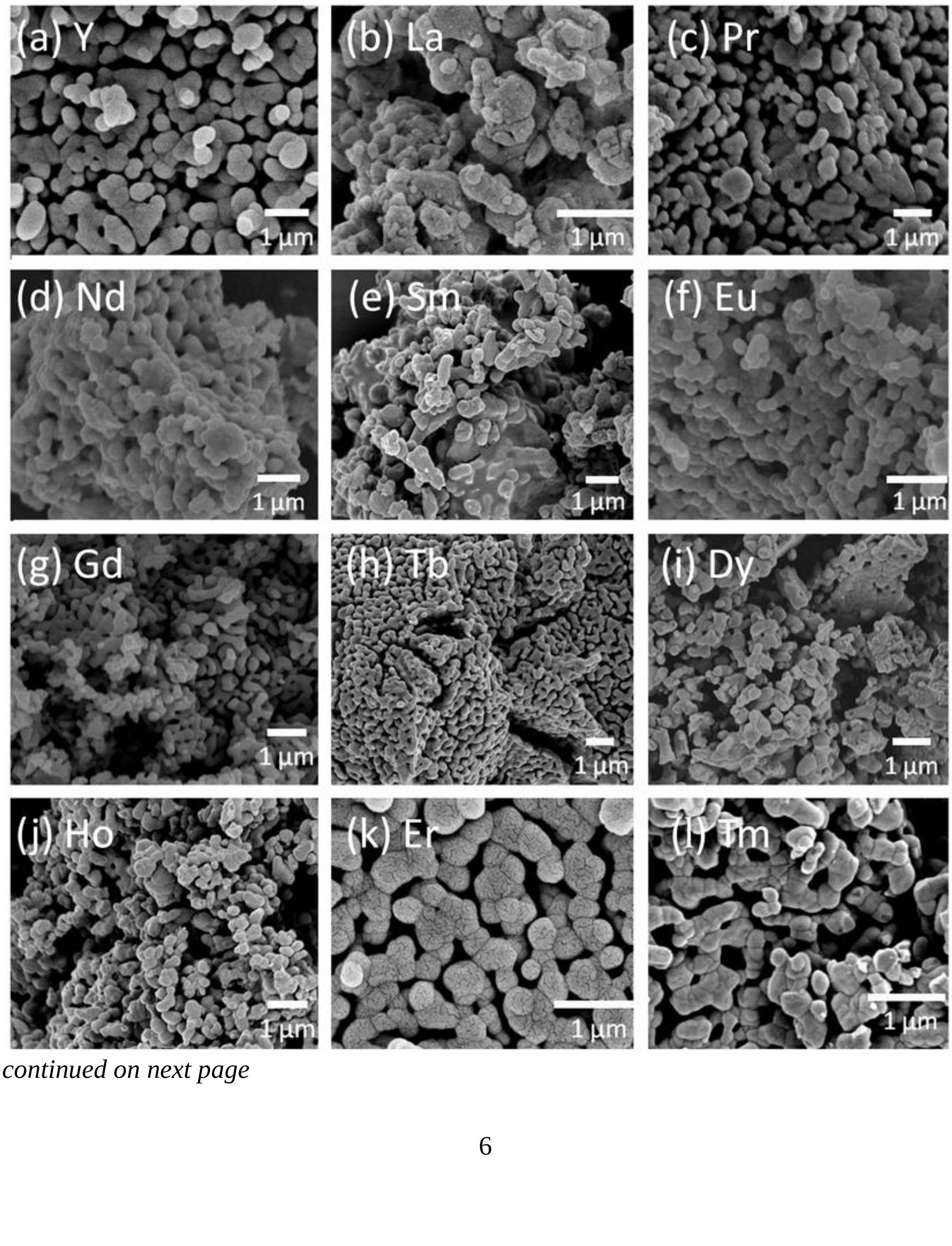

*Figure 6 to be continued on next page*

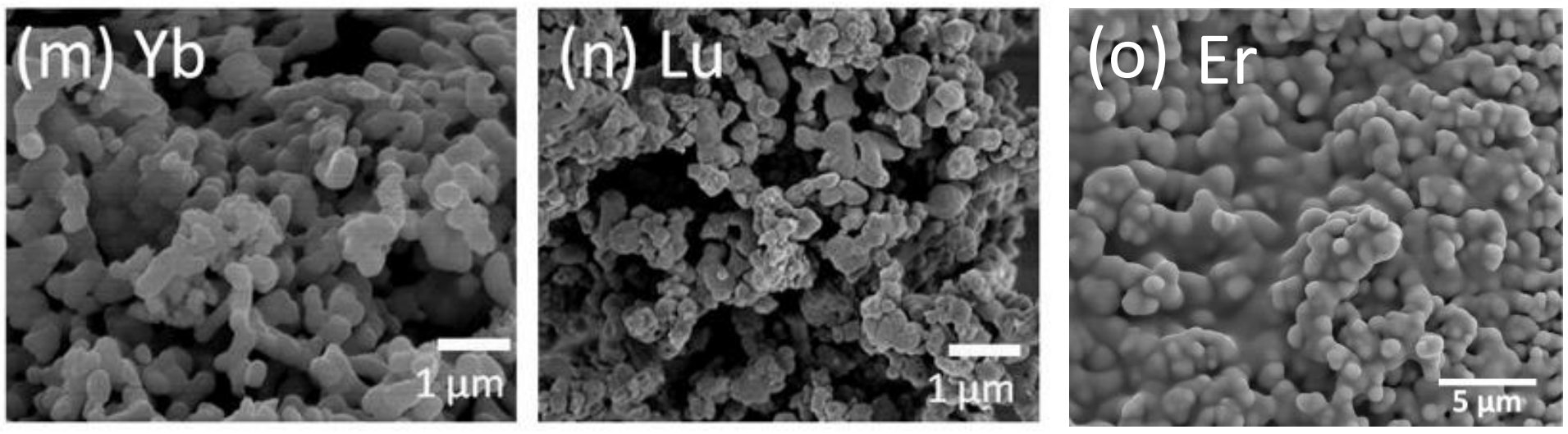

**Fig. 6 (a-n)** SEM micrographs of (RE)$CrO_3$ powders. **(o)** $ErCrO_3$ pellet after densification sintering at 1300° C for 15 hours.

## 3.4 Magnetization measurements

Figure 7 displays a summary of the magnetic data obtained from the (RE)$CrO_3$ series in powder form. In Figure 7a the onset of $Cr^{3+}$-$Cr^{3+}$ ordering can be clearly seen by a sharp increase in magnetization at the Neel transition temperature $T_{N1}$. The representative examples of $GdCrO_3$ and $ErCrO_3$ show no significant differences between ZFC and FC susceptibility curves and $Cr^{3+}$ ferrimagnetic ordering is not indicated. Contrarily, the $PrCrO_3$ sample does show differences between ZFC and FC curves appearing below $T_{N1}$ = 239 K and $Cr^{3+}$ ferrimagnetism is indicated. Such ferrimagnetic ordering is confirmed on this and several other species as indicated in Figure 7b, showing small hysteresis in the magnetization vs applied field curves. The RE-RE ordering temperature $T_{N2}$ for $ErCrO_3$ can be identified in Figure 7a by a peak-like structure in the $\chi$ vs $T$ curve at low temperature. All Curie-temperatures extracted are summarized in Figure 7c. The Figure 7a inset shows the reciprocal magnetic susceptibility $1/\chi$ vs $T$. Above $T_{N1}$, the curves are approximately linear and follow the Curie–Weiss law. The Curie–Weiss fits are indicated by solid lines. From such fits the effective magnetic moments $\mu_{eff}$ and the Weiss constant $\Theta$ can be extracted. All $\mu_{eff}$ values extracted are presented in Figure 7d, together with theoretical values calculated from Landolt-Börnstein tables. The $\Theta$ – values obtained are all negative, which indicates predominantly antiferromagnetic $Cr^{3+}$-$Cr^{3+}$ exchange interactions.

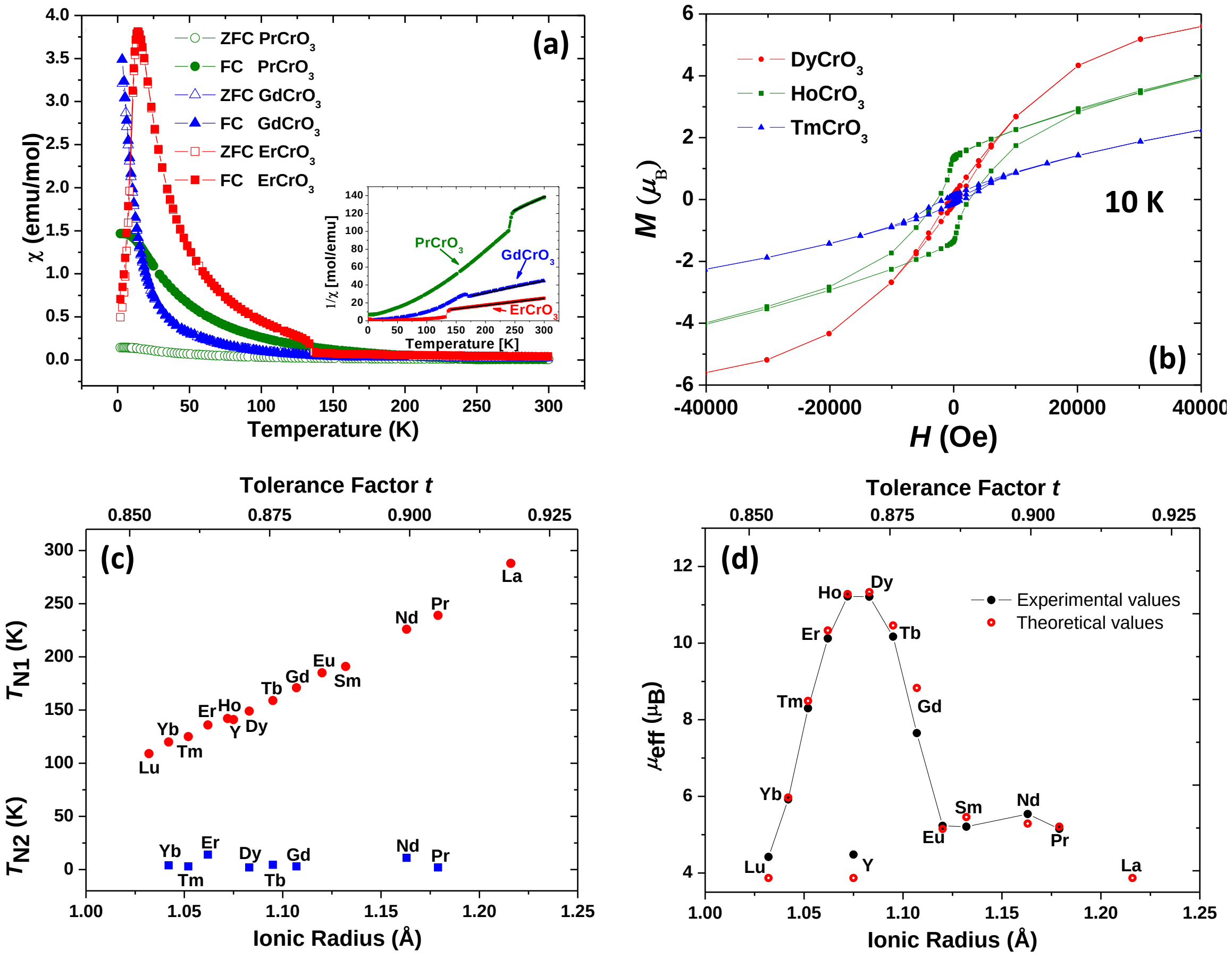

**Fig. 7 (a)** Magnetic susceptibility $\chi$ vs $T$ for ZFC and FC measurements. Inset: $1/\chi$ vs $T$ Curie-Weiss plots **(b)** Magnetization $M$ vs applied magnetic field $H$ at 10 K. **(c)** Neel temperature for $Cr^{3+}$-$Cr^{3+}$ ordering $T_{N1}$ and for RE-RE ordering $T_{N2}$. **(d)** Effective magnetic moments $\mu_{eff}$ vs ionic radius, extracted from the Curie Weiss plots in the Inset of (a). Reproduced from Ref.[6] with permission from the American Chemical Society (ACS © 2013).

## 4. Conclusions

The full (RE)$CrO_3$ series has been synthesized by microwave techniques. SEM analysis reveals a large variety of different shapes of the powders obtained and XRD gives a clear indication of the desired centrosymmetric orthorhombic *Pnma* perovskite phase in all species. The increasing tilting of the perovskite $CrO_6$ octahedra with decreasing size of the RE-cation can be investigated and quantified by Rietveld refinement analysis of XRD pattern. The results were confirmed by atomic resolution HRTEM images, which display the octahedral tilting in real space. The $Cr^{3+}$-$Cr^{3+}$ ordering temperature $T_{N1}$ increases with decreasing tilting angle (Figure 7c), when the Cr-O-Cr angle approaches the ideal 180° which favors magnetic exchange interactions. All such findings validate the novel synthesis route employed. It can therefore be concluded that combined XRD, TEM and SEM techniques are ideal to reveal the formation and display the particle microstructure of a desired phase synthesized by a novel green chemistry route.

**Acknowledgements** Support from the Madrid Community (Materyener S2009/PPQ-1626) and the MICINN/ MINECO (MAT 2007-31034 and MAT 2010-19837-CO6-03) is gratefully acknowledged. R.S. acknowledges a Ramon y Cajal fellowship.